# Periodic feature characterization in nanostructured surfaces and emulsions

André Guerra[1,2,3]*, Ziheng Wang[2], Samuel Mathews[2], Alejandro D. Rey[2], Kevin De France[1†]

[1]Department of Chemical Engineering, Queen's University, Kingston, ON, Canada

[2]Department of Chemical Engineering, McGill University, Montréal, QC, Canada

[3]Department of Chemical and Biomolecular Engineering, Cornell University, Ithaca, NY, USA

*andre.guerra@mail.mcgill.ca;

†kevin.defrance@queensu.ca

## Abstract

Understanding structure-function relationships is essential to advance the manufacturing of next-gen materials with desired properties and functionalities. Precise and rapid measurement of features like wrinkle size, droplet diameter, and surface roughness is essential to establishing such structure-function relationships. To this end, this work developed feature size and surface morphology characterizations through image analysis in Python and validated them with both synthetic and experimental images. Manual measurements of bio-based surfaces resulted in between 3.3% (N=50, visually simple) and 51.2% error (N=100, visually complex) compared to Python analysis results. This analysis was also used to accurately distinguish multiple feature size populations in a given image (which were missed entirely in manual measurements), and to determine the skewness and kurtosis of biological surfaces in a surface roughness map. This work contributes to a larger goal of developing a robust and computationally cheap platform to analyze complex materials to accelerate structure-function discovery.

# 1. Introduction

In standard Fourier Analysis, a function is represented as a combination of periodic functions (e.g., cosine and sine functions) in a finite Fourier Series. This widely used mathematical transform enables breaking up a complex function (or measured signal) into a summation of "simpler" periodic functions. These can be solved individually, and their solutions can be combined to obtain a solution to the original function. A Discrete Fourier Transform (DFT) discretizes a continuous function into equal-sized intervals to facilitate the development of a Fourier Series to represent it. This algorithm is computationally expensive, as a Fourier Series with $N$ terms requires $2N^2$ operations. Cooley and Tukey developed the Fast-Fourier Transform (FFT) algorithm in 1965[1], greatly reducing this computational requirement to only $2Nlog_2N$ operations. **Figure 1** demonstrates how the computational time complexity of DFT calculations compares to that of FFT. The figure inset emphasizes how the DFT algorithm's time complexity diverges from FFT within only a few terms in the series. This discrepancy is significant when dealing with the Fourier Transform of two-dimensional data, such as images, which can contain thousands of pixel values.

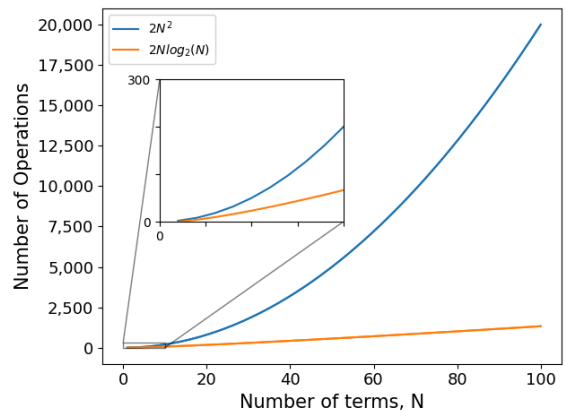

**Figure 1: Time complexity comparison between Fourier Transform algorithms.** This plot demonstrates the number of required operations for the Discrete Fourier Transform (blue) and the Fast-Fourier Transform (orange). The inset demonstrates how quickly the time complexities of these algorithms diverge from each other.

Understanding structure-function relationships of materials is essential to leverage these relationships to manufacture materials with desired properties and functionalities. Bottom-up fabrication of functional materials by mimicking nanostructuring found in Nature has driven various types of bioinspired and biomimetic research. One such domain focuses on utilizing a ubiquitous biological building block - cellulose - as a starting point to develop aligned, porous, and fibrous functional materials. A recent review presents the current knowledge, the state-of-the-art, and the challenges faced in this area[2]. Two-dimensional thin films are of special interest, notably for their biomedical applications such as tissue engineering[3], biosensing and actuating[4,5], or flexible electronics[6,7].

Previous work has been performed to apply FFT analysis to SEM images of two-dimensional nanostructured surfaces for periodic feature identification[8]. Nanowrinkled surfaces were produced by a buckling mechanism[9] applied to a thin film via thermal-induced shrinking of a pre-stressed polystyrene substrate[10,11]. This previous approach utilized the proprietary software MATLAB, making it inaccessible to individuals without an active license. This work develops an open-source Python implementation of image analysis using core packages and standard libraries, broadening the algorithm's reach and allowing for continuous development of a software suite focused on nanostructured surface analysis. The roughness of a 2D nanostructured surface can be quantified by the statistical distribution of its height profile. The third and fourth moments (skewness and kurtosis) describe the bias and magnitude of the peak or valley features on the surface[12,13]. The Python analysis presented in this work also evaluates skewness and kurtosis from atomic force microscopy (AFM) images. The analyses and results presented here may offer complementary information to fiber size analysis offered by the open-source software FiberApp[14].

The objective of this work was to (1) develop periodic feature characterization tools using the Python programming language, (2) validate the tools developed using synthetic images, and (3) apply our approach to experimental images having widely varied morphologies and periodic characteristic features: cellulose-based nanowrinkled surfaces[9–11,15], emulsions[16], and lysozyme amyloid fibrils. This work contributes to a larger goal of developing a robust and computationally cheap platform to analyze complex structures in natural materials to accelerate structure-function discovery and potentiate new ways to achieve biomimicry. The following section presents a detailed background on the image analysis methodology and its validation. All code, validation test cases, and data presented can be found in a GitHub repository[17]. This is followed by a discussion on the results obtained from the tools developed and their validation. The final section presents conclusions drawn, comments on the performance of the characterizations demonstrated, and suggestions for future work. Finally, an appendix (Supplemental Information) provides additional test cases and analyses for reference.

## 2. Methods and Validation

## 2.1. One-Dimensional Signal Analysis

An arbitrary signal can be decomposed into a series of periodic functions, which, when properly combined, accurately represent the original signal. This means that the inverse can be done to make a synthetic signal for testing purposes. **Figure 2** presents two signals (orange and green in the inset plot) combined and introduced to random noise to obtain a noisy raw signal (blue). The noisy raw signal was used in this section as a test case to demonstrate a one-dimensional FFT analysis. Signal 1 ($s_2$) was defined as a 5 Hz sinusoid with an amplitude of magnitude 2, while Signal 2 ($s_2$) was a 20 Hz sinusoid with an amplitude of magnitude 1. Equations 1 and 2 represent the two signals, respectively.

$$s_1 = 2\,sin(2\pi(5t^*)) \tag{1}$$

$$s_2 = sin(2\pi(20t^*)) \tag{2}$$

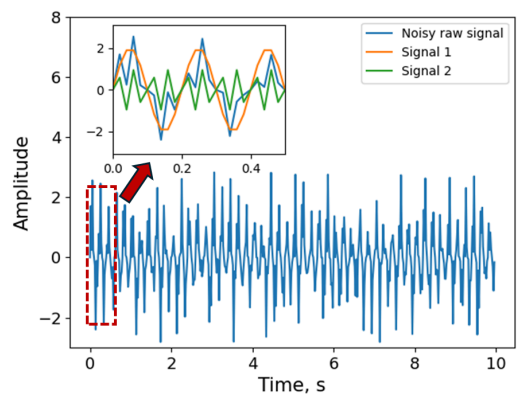

**Figure 2: Decomposition of a noisy signal.** The plot presents a noisy raw signal (blue). The inset plot shows a zoomed section of the raw signal and includes its two components (Signal 1 in orange and Signal 2 in green).

The FFT analysis used the *numpy* Python package functions to conduct all transformations and manipulations. **Figure 3** demonstrates the result of the analysis of the noisy raw signal introduced in **Figure 2**. The procedure followed was (1) the noisy raw signal was transformed using the *numpy* FFT algorithm, (2) the transformed result was then center-shifted to emphasize the symmetry around frequency zero (**Figure 3b**), (3) the squared magnitude of the FFT result provides the power spectral density (PSD) of the signal, which is finally (4) normalized to absolute intensity units by dividing all PSD values by the maximum intensity peak value (**Figure 3c**). The PSD of the signal reveals dominant frequencies hidden in the noisy raw signal - a 5 Hz signal and a 20 Hz signal (**Figure 3c**). Additionally, the size of the corresponding peaks in panel (b) indicates that the 5 Hz signal is roughly twice the intensity of the 20 Hz signal. These are the frequencies and relative amplitudes of Signals 1 and 2 (**Eqs. 1-2**), which were used to generate the noisy raw signal described above. The added noise causes some deviation from an exact 2:1 ratio.

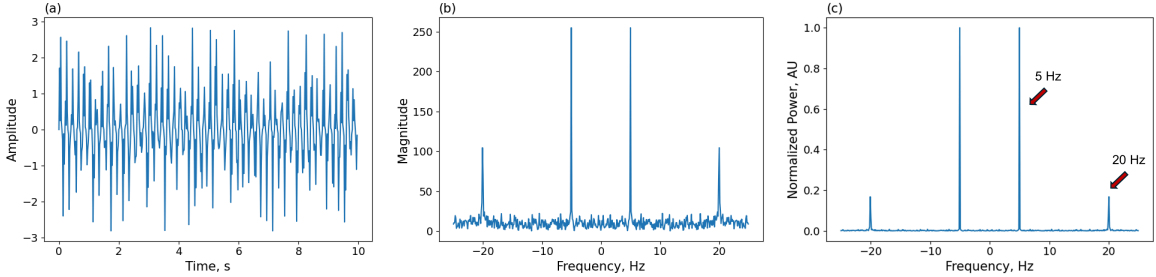

**Figure 3: FFT analysis and results.** (a) The original noisy raw signal, (b) the magnitude of the center-shifted FFT of the signal, and (c) the normalized power spectral density (PSD).

## 2.2. Two-Dimensional Signal Analysis (Images)

The one-dimensional analysis described above can be extended to two-dimensional signals, such as a grayscale or RGB array defining each pixel value of an image. To test and validate the two-dimensional analysis developed in Python, this work uses four different synthetically generated images inspired by previous work[8] - an original image of vertical lines (stripes), and increasingly complex patterned images made by modifying the original image. These include a chevron pattern (zig-zags), a tessellation of sections of the original image, and tessellations with increased fragmentation. These patterned images were created to simulate surfaces with various morphologies, whose structure can be characterized using the 2D signal analysis presented here.

### 2.2.1. Vertical Lines

The first test case developed was an image with vertical black and white lines (stripes). An 8-bit image was generated as a two-dimensional *numpy* array with 900 elements in each dimension and populated with zeros. Zero values are interpreted as black pixels in an 8-bit image. Stripes were defined as 50 pixels in width, and their values in the array were updated to 255, interpreted as white. **Figure 4** presents the original image (panel a) and the two-dimensional FFT (2D-FFT) analysis performed to extract periodic features of this image. **Figure 4b** presents the image's two-dimensional power spectrum density (2D-PSD), obtained by calculating the magnitude of the center-shifted 2D-FFT of the image array. The 2D-PSD of the image contains information about the frequencies of the signals that make up the original image. However, it is difficult to evaluate it in this form. The center-shifted 2D-PSD presents the zero frequency in the center position of the 2D array and increases radially outward. To transform this 2D-PSD into a one-dimensional spectral density, the radial average of the 2D-PSD was determined (**Figure 4c**). The radially averaged PSD enables the identification of the frequency of periodic signals in the original image. Finally, the radially averaged PSD frequencies are transformed from pixels to the appropriate length scale. In the test case of the vertical lines, the pixels were arbitrarily taken to represent 1 µ$m$. **Figure 4d** presents the final radially averaged PSD, and the most prominent frequency (*f*) peak was identified as 0.01 µm$^{-1}$, equivalent to a period (*T*) of 100 pixels (**Eqs. 3-4**). Recalling that the original image was designed with black and white stripes 50 pixels in length, the period of 100 pixels refers to one cycle in the vertical lines pattern (two stripes - one black and one white). This result validates the 2D-FFT analysis on the vertical lines test case, as the periodic feature identified corresponds to the periodic design of the generated image. It should be noted that smaller peaks are detected at 0.03, 0.05, and 0.07 µm$^{-1}$. These are much smaller in intensity and represent auxiliary periodicity in the image as multiples of the fundamental frequency (0.01 µm$^{-1}$).

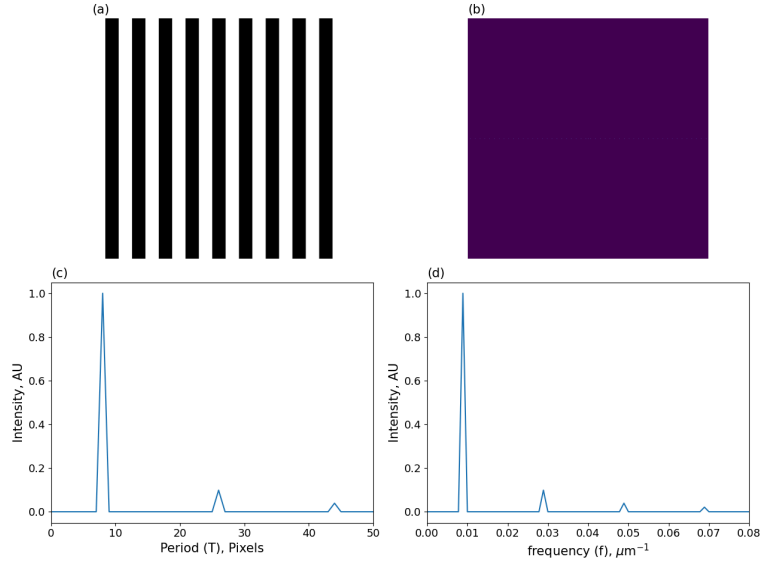

**Figure 4: Vertical lines.** (a) The 900x900 pixel 8-bit image of black and white stripes 50 pixels in width, (b) the natural log of the magnitude of the center-shifted 2D-FFT of the image (2D-PSD) of the image), (c) the radially averaged 2D-PSD of the image in the pixel frequency domain, and (d) the radially averaged 2D-PSD of the image in the spatial frequency domain.

$$f = 0.01\ \mu m^{-1} \tag{3}$$

$$T = \frac{1}{0.01\ \mu m^{-1}} = 100\ \mu m(\frac{1\ pixel}{1\ \mu m}) = 100\ pixels \tag{4}$$

## 2.2.2. Increasingly Complex Patterned Images

The same 2D-FFT analysis on the vertical lines test case was applied to other synthetically generated images with increasingly more complex patterns. All test case images were created by modifying the original image of the vertical lines described above. The analyses of these test cases are presented in this section, with one additional case (chevron) included in the Supplemental Information section (**Figure S1**). The chevron test case features a zig-zag pattern (**Figure S1a**) generated by periodically shifting pixels left and right to create 45-degree zig-zags. The radially averaged PSD identifies the same dominant frequency (0.01 μm$^{-1}$) as in the original image, corresponding to the period of 100 pixels (**Figure S1d**). The chevron pattern does not introduce any significant new signal to the original image and, thus, does not change its most prominent periodic feature.

Noise in a signal can be interpreted as heterogeneity - it introduces randomness to the underlying signal. Higher levels of noise result in a broader primary peak and smaller secondary peaks in the signal's radially averaged power spectrum density. In other words, the identification of noise in the signal corresponding to the main periodic feature of the image indicates heterogeneity in the physical pattern that the image captures (e.g., surface wrinkles). The width of noisy signals can be quantified using the full-width at half-maximum (FWHM). This is the width of the peak at the intensity height that is half of the maximum (peak) intensity (**Eq. 5a**). Additionally, the peak quality factor (PQF) can also be quantified as the ratio between the peak's maximum intensity of the power spectrum (*S*) and its FWHM (**Eq. 6**). The frequencies $S(f_1)$ and $S(f_2)$ are the coordinates where the intensities are half of the peak's maximum intensity ($S(f_0)$) (**Eq. 5b**).

$$FWHM = f_2 - f_1 \tag{5a}$$

$$S(f_2) = S(f_1) = \frac{1}{2}S(f_0) \tag{5b}$$

$$PQF = \frac{f_0}{FWHM} \tag{6}$$

The following two test cases introduce non-negligible noise to the initial image by random placement and orientation of tiles in a tessellation. First, the tessellation was generated using the photo editor software Photopea[18] to create tiles from the original image of the vertical lines and to arrange these tiles to form the pattern (**Figure 5a**). Next, a fragmented tessellation was created by generating new tiles from the tessellation in **Figure 5a** to create a new image with higher randomness (or noise) (**Figure 6a**). These images were analyzed using the same procedure described above to identify periodic features encoded in the pixels of the image. Since the images were created from the same original pattern (black and white vertical stripes of 50 pixels in width), the most prominent periodic feature identified by the radially averaged PSD was the same 100-pixel period as in the first two test cases. Their FWHM and PQF were also calculated to quantify each image's heterogeneity. It is evident from **Figures 5d** and **6d**, that the fragmented tessellation has a lower FWMH and higher PQF than the simple tessellation. This may seem counterintuitive at first. However, the heterogeneity in **Figure 5a** is higher than in **Figure 6a** because it contains large pieces with different orientations. In contrast, the pieces in the fragmented tessellation are smaller and their orientation is even more random, so the overall image approaches a new averaged signal. This can be analogized to a mixture of two pure components. Initially, the mixture is heterogeneous (low entropy of mixing), but as it is continuously mixed, it takes on an average concentration between the two components (high entropy of mixing). These test cases use FWHM and PQF to interpret noise in the periodic features in an image and, thus, demonstrate the potential to examine the periodic features and structure of a physical system (e.g., surfaces or emulsion droplets) from its image.

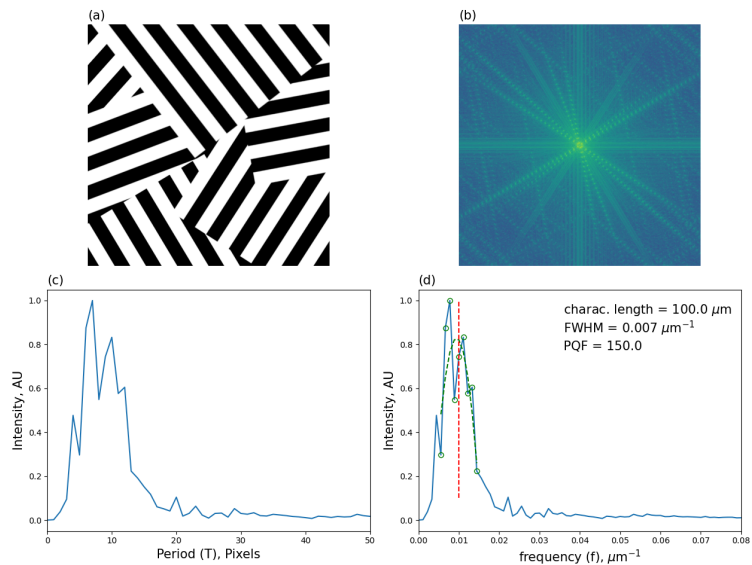

**Figure 5: Tessellation of the image of the vertical lines.** (a) The 900x900 pixel 8-bit image of black and white stripes 50 pixels in width was used to create tiles to form this tessellation image, (b) the magnitude of the center-shifted 2D-FFT of the image (i.e., 2D-PSD of the image), and (c) the radially averaged 2D-PSD of the image in the pixel frequency domain, and (d) the radially averaged 2D-PSD of the image in the spatial frequency domain. The characteristic length, FWHM, and PQF are presented in panel (d).

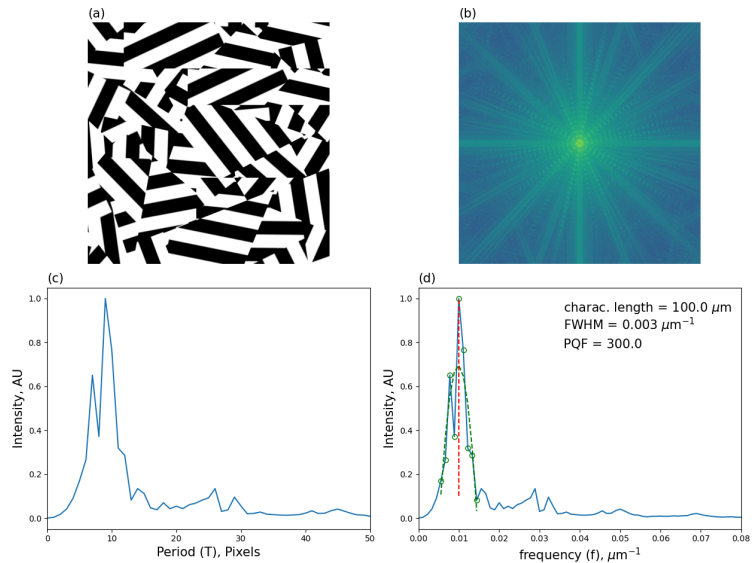

**Figure 6: Fragmented tessellation image of the vertical lines.** (a) The 900x900 pixel 8-bit image of a black and white tessellation image presented in **Figure 5a** was used to create tiles to generate a new tessellation with a higher degree of fragmentation, (b) the magnitude of the center-shifted 2D-FFT of the image (i.e., 2D-PSD of the image), and (c) the radially averaged 2D-PSD of the image in the pixel frequency domain, and (d) the radially averaged 2D-PSD of the image in the spatial frequency domain. The characteristic length, FWHM, and PQF are presented in panel (d).

## 2.3. Surface Roughness Determination

A distribution function's first and second moments are commonly known as the *mean* and *variance* (i.e., root mean square), respectively. The skewness is the third moment, a measure of bias of the mean, median, and mode to one side (left or right skew) in the distribution. Kurtosis is the fourth moment, which measures the heaviness of the distribution's tail - how much of the distribution populates the left and right tails. In surface morphology, a surface's height distribution statistics can be used to describe and characterize its nanoscale periodic features. A recent review presents a detailed description of pattern formation and characterization in wrinkled surfaces of liquid crystal materials using the surface's height distribution probability density function (PDF)[12]. The skewness and kurtosis of the surface are referred to as surface roughness parameters.

Atomic force microscopy (AFM) produces the PDF of surface profiles as an output image. The surface's PDF can then be used to compute its third and fourth moments - skewness ($Rsk$) and kurtosis ($Rku$). A normally distributed PDF has $Rsk = 0$ and $Rku = 3$. When interpreting these parameters as indicators of surface roughness, progressively larger absolute values of $Rsk$ are associated with increasing deviation from normality. Negative $Rsk$ values indicate low valleys (**Figure 7a**), and positive $Rsk$ values indicate tall peaks (**Figure 7b**). Kurtosis ($Rku$) describes the proportion of the surface comprised of peaks and valleys. A high kurtosis is associated with sharp protrusions or valleys (**Figure 7b**), while a low kurtosis indicates surfaces with smoother undulations (**Figure 7c and d**).

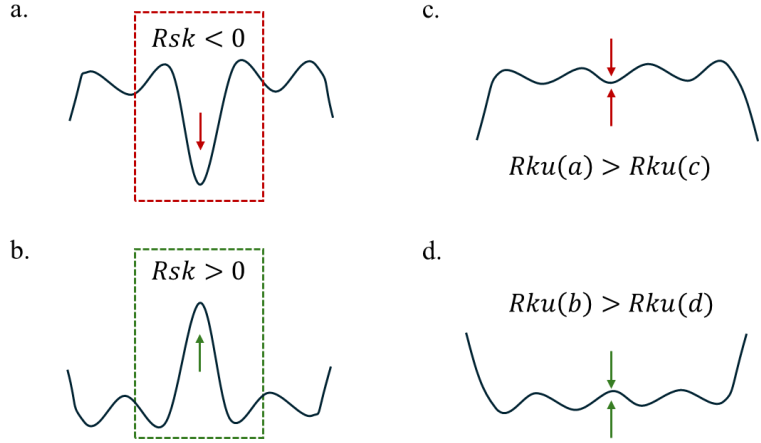

**Figure 7: Surface roughness parameters.** Sample surface height profiles where (a) valleys are prevalent ($Rsk < 0$) and progressively deeper as $Rsk$ decreases, (b) peaks are prevalent ($Rsk > 0$) and progressively sharper as $Rsk$ increases, (c) and (d) kurtosis values converging to $Rku = 3$, indicating a smoother undulating surfaces compared to deeper and sharper features in (a) and (b).

The root mean square, skewness, and kurtosis of all surface height profiles examined in this work were computed through two-dimensional (2D) integrals on the surface height profile ($h$) as presented in **Eqs. 7-9**. These computations are conducted (i) with dimensionless variables (*) and (ii) using area finite elements ($dA$). The dimensionless variables are necessary as skewness and kurtosis are dimensionless moments, but also enable direct comparison of skewness and kurtosis of surfaces of different scales (e.g., airplane wings and butterfly wings). The area integration contrasts with alternative calculation methods in which a double integral may be integrated sequentially as two single integrals, one in $dx$ and another in $dy$. Although mathematically equivalent, in the context of surface properties, this results in a loss of information. For surface profile skewness and kurtosis to be physically meaningful indications of surface roughness, the computed moments must retain information about the proximal points on the curved surface. In other words, at any given point on the surface profile, geometric surface properties such roughness are directly affected by that point's height value and the height values of points in proximal regions (e.g., nearest-neighbors, next-nearest-neighbors, etc.)[19]. The 2D integration with an area finite element ($dA$) enables the retention of this information in the resultant moments. Additionally, integration in $dA$ enables any parametrization of the surface without losing the generality of **Eqs. 7-9**.

Root mean square (2nd moment) $$Sq = \sqrt{\frac{1}{A^*}\iint_\Omega h^{*2} dA^*} \qquad (7)$$

Skewness (3rd moment) $$Rsk = \frac{1}{A^* Sq^{*3}}\iint_\Omega h^{*3} dA^* \qquad (8)$$

Kurtosis (4th moment) $$Rku = \frac{1}{A^* Sq^{*4}}\iint_\Omega h^{*4} dA^* \qquad (9)$$

# 3. Results and Discussion
## 3.1. Biaxial Nanowrinkled CNC-POEGMA Thin Films

The Python surface analysis developed in this work was applied to scanning electron microscopy (SEM) images of biaxially buckled thin films composed of cellulose nanocrystals (CNCs) and poly (oligoethylene glycol methacrylate) (POEGMA). Details on the manufacturing process of these thin films, including solution preparations, film deposition, and the thermal buckling procedure to create nanoscale wrinkles, have been previously described elsewhere[10]. **Figure 8** presents the procedure and results from the Python surface analysis developed here. The original image (**Figure 8a**) was pre-processed by low band-pass filtering to remove low-frequency anomalies using a difference of Gaussians (**Figure 8b**). The filtered image was processed using the Canny edge detection algorithm[20] to identify the boundaries of nanowrinkles (**Figure 8c**), and the edge-detected image had its 2D-PSD calculated (**Figure 8d**). The 2D-PSD was radially averaged and converted to the spatial domain to obtain the frequency of periodic features (wrinkles) on the surface (**Figure 8e**). It includes a quadratic regression over the most prominent peak to identify the size of the most common wrinkle.

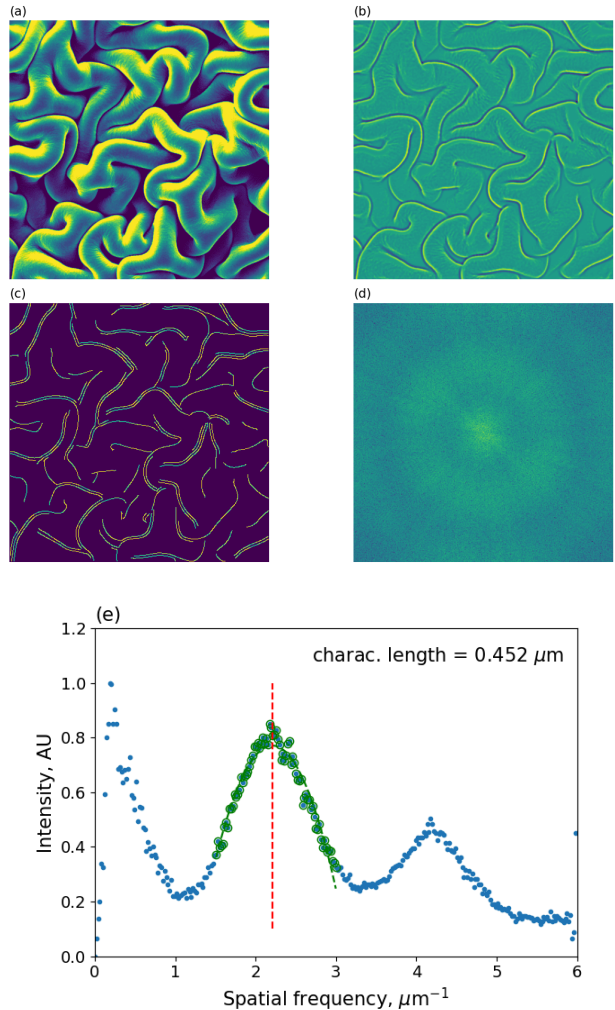

**Figure 8: Biaxial wrinkles in CNC-PEOGMA thin films.** (a) The original SEM image collected by De France et al.[10], (b) low band-pass filtered image by a difference of Gaussians ($\sigma_{Gauss} = 1.1$), (c) edge-detected image using the Canny algorithm ($\sigma_{Canny} = 1.4$), (d) 2D-PSD of the edge-detected image, and (e) the radially averaged power spectral density with quadratic regression applied to the most prominent peak in the scale of interest to identify characteristic length of periodic features of the surface (biaxial wrinkles) identified by a dashed red vertical line.

The final output from the Python analysis for the biaxially wrinkled CNC-POEGMA surface indicated that its characteristic length - the most commonly occurring wrinkle size - was 0.452 µm. To validate this result, the same image was manually analyzed using Fiji[21] (an open-source ImageJ software with several libraries for added analytical capabilities). The SEM image's scale bar (20 µm) was used to calibrate the measure feature in ImageJ. Fifty ($n = 50$) manual measurements were made by drawing line segments between the edges of wrinkles in the image. Since this surface was biaxially wrinkled, the measurements were performed randomly (in all directions and all positions) on the image. This resulted in a measured wrinkle size of 0.536 µm, corresponding to an 18.8% error relative to the Python analysis estimate. The manual measurements were repeated with a larger sample size ($n = 100$), which improved the wrinkle size estimate to 0.492 µm (8.9% error).

## 3.2. Uniaxial Nanowrinkled CNC-POEGMA Thin Films

The original experimental work demonstrating the manufacturing process for the CNC-POEGMA thin films analyzed above also produced uniaxially wrinkled surfaces. The uniaxial wrinkling was achieved by constraining one axis direction while the other was allowed to undergo thermal buckling[10]. The analysis above was repeated for a uniaxially wrinkled CNC-POEGMA thin film surface, presented in **Figure 9**. The Python analysis indicated a 0.44 µm characteristic length for the uniaxial wrinkles (**Figure 9e**). Manual measurements of the wrinkles were performed using Fiji software to validate this estimation, as described above. A sample of 50 measurements ($n = 50$) was collected by drawing line segments between wrinkle edges at random placements on the image. The direction of the line segments was preferentially taken to be perpendicular to the wrinkles observed in the SEM image. This resulted in a measured wrinkle size of 0.426 µm, corresponding to a 3.3% error relative to the Python analysis estimate. The manual measurement of the uniaxial wrinkles resulted in a considerably lower error, even with a smaller sample size ($n = 50$). We attribute this to the uniaxial wrinkles having increased uniformity/homogeneity, which facilitated their manual measurement. Additionally, the uniaxial nature of the morphology eliminated one variability in manual measurements - the direction. In this case, the consistent manual measurements perpendicular to the wrinkled motif enable increased consistency in manual measurements. In contrast, the biaxial manual measurements described in the previous section suffered in part from higher variability due to the larger number of possible directions for the measurement. In the next section, an experimental test case with higher heterogeneity in its morphology will be contrasted with the results presented so far.

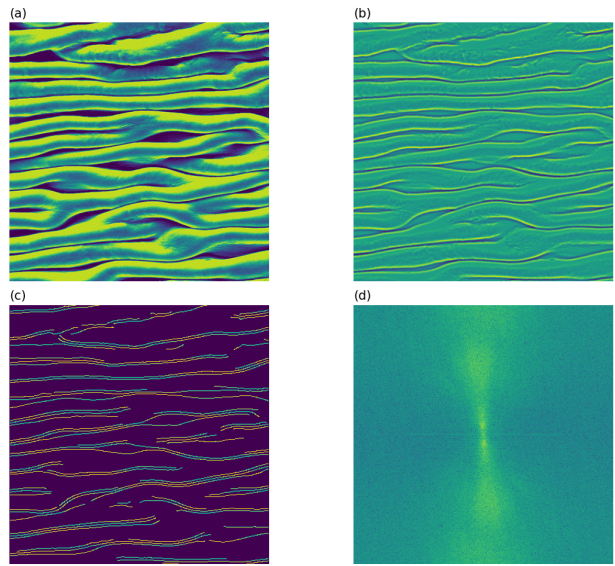

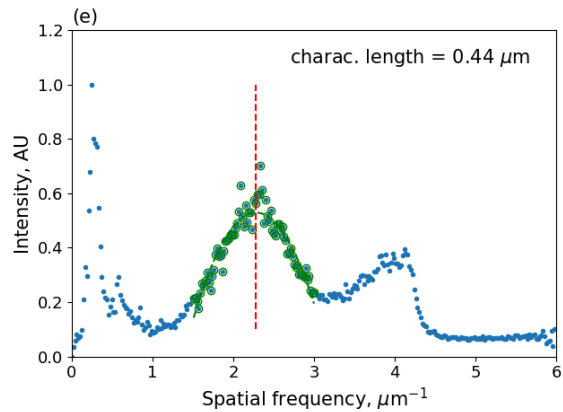

**Figure 9: Uniaxial wrinkles in CNC-PEOGMA thin films.** (a) The original SEM image collected by De France et al.[10], (b) low band-pass ($\sigma_{low} = 1.0$) and high-band pass ($\sigma_{high} = 10$) filtered image by a difference of Gaussian, (c) edge detected image using the Canny algorithm ($\sigma_{Canny} = 2.0$), (d) 2D-PSD of the edge-detected image, and (e) the radially averaged power spectral density with quadratic regression applied to two prominent peaks in the scale of interest to identify characteristic length of periodic features of the surface (uniaxial wrinkles) identified by a dashed red vertical line.

## 3.3. Biaxial Nanowrinkled CNC-XG Thin Films

An additional test of the Python wrinkle size analysis was performed using a biaxially wrinkled thin film composed of CNCs and xyloglucan (XG). This nanostructured surface was selected as it exhibits a different wrinkle morphology than the previously analyzed surfaces. **Figure 10** presents the same analysis procedure described in the previous section, applied to the CNC-XG nanowrinkled surface. It is evident from **Figure 10b** that the surface exhibits increased variability in the morphology of the wrinkles formed. The low and high-band pass filtering improves wrinkle detection by reducing the signals related to the deepest valleys and highest peaks on the surface, which smooths the image and improves the Canny edge detection step (**Figure 10c**). The Python analysis identified the most common wrinkle size to be approximately 1.1 μm. Manual measurements were obtained using ImageJ, as discussed above. A sample of 100 manual measurements ($n = 100$) led to a wrinkle size of 0.535 μm, corresponding to a 51.2% error relative to the Python analysis estimate. We attribute this to variability in wrinkle morphology dramatically reducing the accuracy of manual measurements, even with a larger sample size.

Upon further scrutiny, **Figure 10e** presents a broader distribution of spatial frequencies compared to the easily identifiable peaks in **Figures 6e** and **8e**. This indicates that the CNC-XG surface contains wrinkles of various sizes (nonuniform, in addition to biaxial wrinkling). Moreover, one can identify a second spatial frequency signal near 1.8 μm$^{-1}$ (dashed red box in **Figure 10e**). This secondary signal identifies a population of smaller wrinkles (equivalent to 0.556 μm), in better agreement with the manual measurement of 0.535 μm. The manual measurements were therefore likely biased toward this second population of wrinkles. This can be either because these are easier to visually detect or due to a subconscious selection bias for smaller features. Smaller features are likely interpreted as "nanowrinkles," while larger features are dismissed as "gaps."

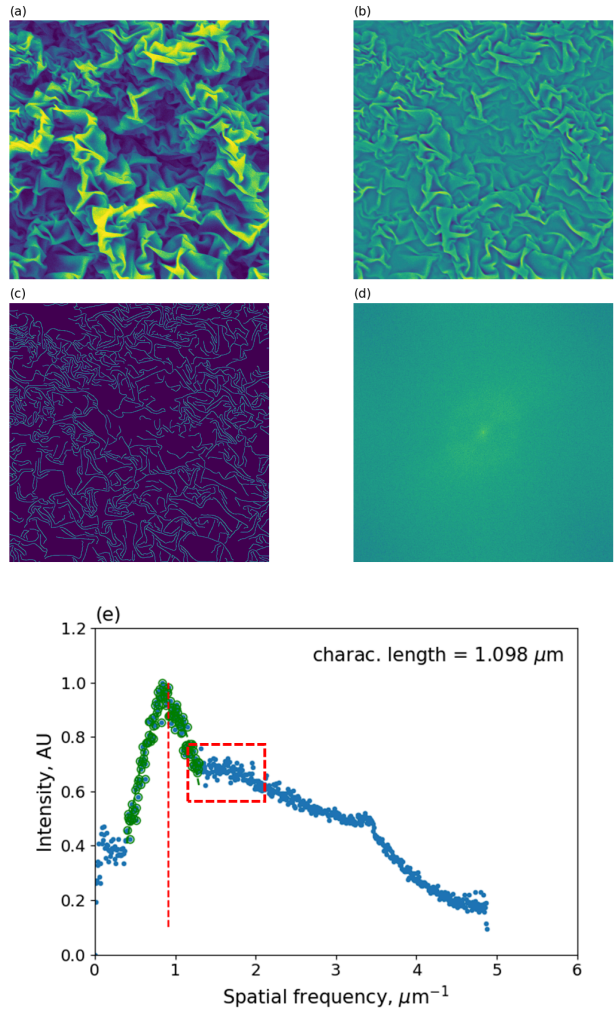

**Figure 10: Biaxial wrinkles in CNC-XG thin films.** (a) The original SEM image of the nanostructured surface, (b) low band-pass ($\sigma_{low} = 3.0$) and high-band pass ($\sigma_{high} = 12$) filtered image by a difference of Gaussian, (c) edge detected image using the Canny algorithm ($\sigma_{Canny} = 1.4$), (d) 2D-PSD of the edge-detected image, and (e) the radially averaged power spectral density with quadratic regression applied to the most prominent peak in the scale of interest to identify characteristic length of periodic features of the surface (biaxial wrinkles) identified by a dashed red vertical line.

## 3.4. CNC Emulsion Droplets

In addition to nanowrinkled surfaces, this work also tested the Python analysis discussed in the previous section on CNC-based emulsions. The periodic feature identification and characteristic length determination were used in this case to quantify droplet population diameters. **Figure 11** presents the results of this analysis. As with nanowrinkled surfaces, the emulsion images were filtered through Gaussian differences, and Canny edge detection was used to identify droplet boundaries. The 2D-FFT was used to derive the 2D-PSD, and the two largest droplet populations were identified (**Figure 11e-f**). The characteristic lengths (diameters) were determined to be 6.1 and 0.87 µm. Manual measurements were performed using ImageJ as described above. In this case, the manual sampling of droplet diameters was more difficult due to the number of droplets and their variety in sizes. It was exceptionally challenging to distinguish between population sizes by eye. The manual measurements resulted in an average droplet diameter of 2.1 µm, which lies between the two population sizes identified by the Python analysis. Two additional images of the same system were analyzed for comparison, and the results were included in the Support Information. The characteristic lengths obtained agreed with the values from **Figure 11** - 8.9 and 0.89 μm (**Figure S2**) and 6.2 and 0.86 μm (**Figure S3**).

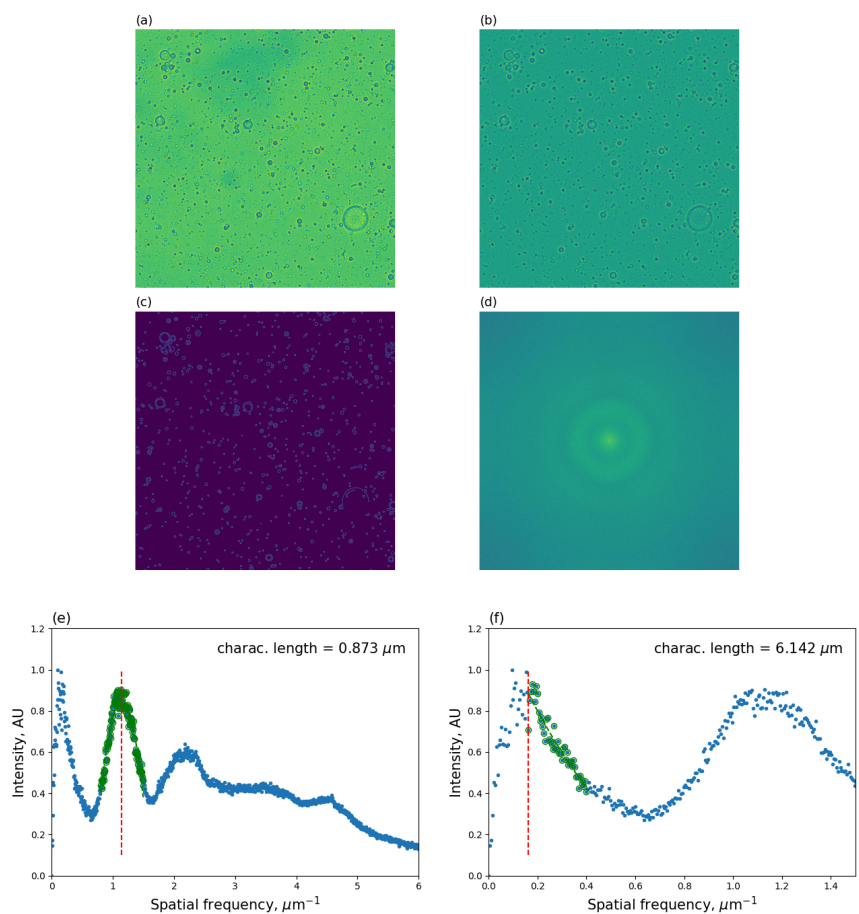

**Figure 11: CNC emulsion with various-sized droplets entrained.** (a) The original image of the CNC emulsion with droplets, (b) low band-pass ($\sigma_{low} = 3.0$) and high-band pass ($\sigma_{high} = 12$) filtered image by a difference of Gaussian, (c) edge detected image using the Canny algorithm ($\sigma_{Canny} = 1.4$), (d) 2D-PSD of the edge-detected image, and (e and f) the radially averaged power spectral density with quadratic regression applied to the two most prominent peaks in the scale of interest to identify characteristic length of periodic features of the surface (droplet diameters) identified by a dashed red vertical line.

## 3.5. Characteristic Frequency Noise and Physical Structure

As discussed above, noise in the characteristic frequency signal from the PSD of a surface's image can provide insight into its physical periodic structure. The PSD of nanostructured surfaces and CNC emulsions presented in the sections above were analyzed using the FWHM and PQF defined in **Eqs. 5-6** and demonstrated in **Figures 5** and **6**. This data is presented in **Table 1**. The biaxial and uniaxial CNC-POEGMA surfaces have similar PQF but different FWHM. The biaxial surface has a larger FWHM, which indicates higher noise in the signal frequency that characterizes its periodic structure over that of the uniaxial surface. This is expected as the uniaxial structure changes mainly in one direction, while comparatively, the biaxial surface has a greater directional randomness (noise) in its morphology. Furthermore, comparing the biaxial CNC-POEGMA and CNC-XG surfaces also offers valuable insight into differences in their biaxial structures. Although both surfaces were biaxially wrinkled, the materials exhibit visually different morphologies - the POEGMA surface has rounded wrinkles with more consistent width, while the XG surface has sharper wrinkles of varying sizes. The power spectral density peak analysis indicates that the CNC-XG surface (FWHM=0.89; PQF=1.12) has a comparatively lower signal noise (narrower peak) than the CNC-POEGMA surface (FWHM=1.19; PQF=0.71). This result recalls the tessellation and fragmented tessellation validation test cases discussed above. Here, the CNC-XG surface morphology is highly variable (i.e., fragmented), again analogous to a "better mixed" system with higher entropy, exhibiting a narrower and stronger characteristic peak (lower FWHM and higher PQF). Finally, the FWHM and PQF of the CNC emulsion systems were also presented in **Table 1**. For all three images of the CNC emulsions analyzed, the signals' FHWM and PQF are quite similar, which is expected as the images are different captures of the same sample. Although the droplets captured in the three images are different, the emulsion's droplet characteristic lengths are expected to be similar. As a result, in this case, the FWHM and PQF analysis is useful in establishing consistency between samples of the same population for confidence in their characteristic lengths (droplet diameter) presented above.

**Table 1: Full width at half maximum (FWHM) and peak quality factor (PQF)**. The FWHM quantifies the width of the frequency peak in the power spectral density, while the PQF is the ratio between the peak's maximum value ($f_0$) and its FWHM.

| Case | Figure | FWHM [$\mu m^{-1}$] | PQF |
| --- | --- | --- | --- |
| Biaxial CNC-POEGMA | 8 | 1.19 | 0.71 |
| Uniaxial CNC-POEGMA | 9 | 0.95 | 0.74 |
| Biaxial CNC-XG | 10 | 0.887 | 1.12 |
| CNC Droplets (1) | 11 | 0.688 | 1.31 |
| CNC Droplets (2) | S2 | 0.619 | 1.55 |
| CNC Droplets (3) | S3 | 0.619 | 1.10 |

## 3.6. Surface Roughness Estimation of Lysozyme Amyloid Fibrils

This work implemented a surface roughness analysis based on the statistical distribution of the height profile of nanofibers deposited on a substrate. Suspensions of hen egg white lysozyme (HEWL) amyloid fibrils (AF) were produced, and their surface profiles were determined via AFM. The HEWL AF suspensions were synthesized by incubating HEWL (2 wt.%) at 90 °C for 24 hours at pH 2 and agitated by constant mixing, as described elsewhere[22,23]. This suspension was spin-coated on freshly cleaved mica substrates whose surface height profile was characterized by AFM (**Figure 12a**). The original HEWL AF sample suspension was divided into two samples: the first was passed through a dialysis membrane (100 kDa MWCO) to remove small/low molecular weight fibers, and the second was sonicated to break up fibers into smaller segments presented in **Figures 10b** and **10c**, respectively.

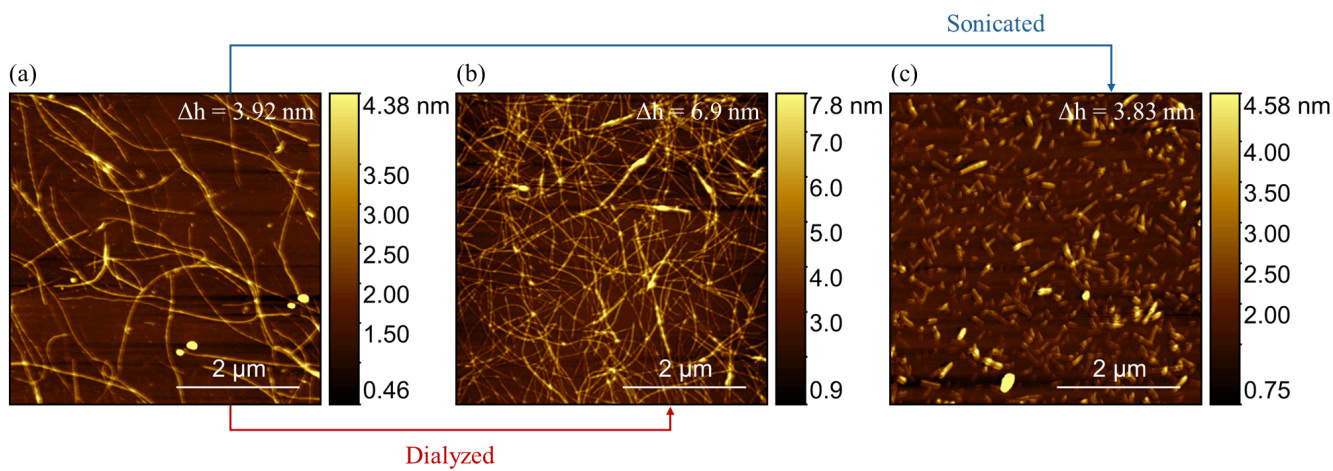

**Figure 12: HEWL AFs.** AFM images (surface height) of HEWL AF synthesized by different protocols for size control. (a) HEWL was incubated at 90 °C for 24 hours at pH 2 and agitated by constant mixing, (b) HEWL AF suspension was passed through a 100 kDa MWCO dialysis membrane to remove smaller (lower MW) fragments, and (c) HEWL AF suspension was sonicated to break up the AFs into smaller fiber lengths.

The AFM images provide insight into the HEWL AF nanofibers. It is evident from **Figure 12** that once the sample is filtered to remove smaller fibers, the resultant surface has a larger spread ($\Delta h = 6.9$ nm) between its highest and lowest points (**Figure 12b**) than that from an unfiltered sample ($\Delta h = 3.92$ nm) (**Figure 12a**). Moreover, the sonicated sample has a lower spread ($\Delta h = 3.83$ nm) (**Figure 12c**). This information from the AFM images may conclude that the sample with the highest "surface roughness" was made from the dialysis membrane filtered HEWL AF sample, while the sonicated sample led to the lowest "surface roughness". However, those are measures of the extremes on the height profile of the surface. Surface roughness is more rigorously characterized by using the moments of the surface's height profile probability distribution functions as described in **section 2.3**. **Figure 11** presents a surface roughness map describing several sample surface height distributions' third and fourth moments (skewness and kurtosis).

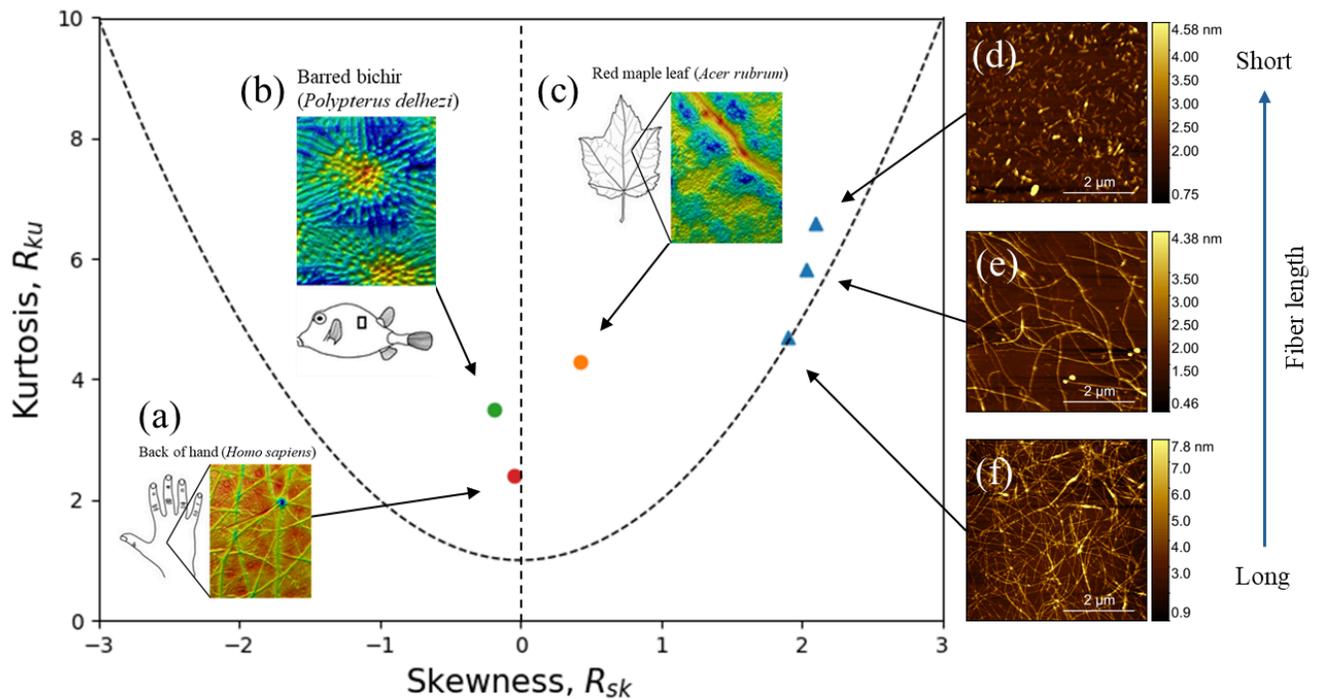

**Figure 13: Surface roughness map.** Sample biological surfaces found in Nature adapted from [24]: (a) back of human hand (*homo sapiens*), (b) barred bichir (*polypterus delhezi*), (c) red maple leaf (*acer rubrum*). HEWL AFs: (d) suspension sonicated to break up AF fibers into smaller length fragments, (e) the original HEWL AF sample, and (f) dialysis membrane (100 kDa MWCO) filtered sample to remove smaller (lower MW) fibers. Fiber length between (e-f) samples is qualitatively indicated on the right scale bar. The dashed black parabola represents the minimum values below which are mathematically impossible according to Peason's inequality $Rsk^2 - Rsk + 1 \leq 0$.

As discussed in **section 2.3**, surface roughness at any given point is directly influenced by its proximal region - its nearest-neighbour and next-nearest-neighbour points. To capture this influence and preserve this information in the surface roughness parameters computed in this work, the integrals in **Eqs. 7-9** were computed using finite area elements. **Figure 13** helps to clarify the differences in surface roughness from the samples presented in **Figure 12** by quantifying more surface characteristics than just surface height difference at extremes. Three sample surfaces were presented for reference (**Figure 13a-c**)[24]. It is noteworthy that naturally occurring biological surfaces tend to have low skewness (Rsk near zero) - these exhibit regular/unbiased roughness. Conversely, the synthetically produced HEWL AF surfaces (**Figure 13d-f**) exhibit irregular roughness ($Rsk \approx 2$). Additionally, the samples of varying fiber length have similar skewness, but their kurtosis differentiates them. As AF fiber length decreases, the surface's kurtosis increases substantially from 4.69 (**Figure 13f**) to 6.6 (**Figure 13d**). This indicates progressively sharper surface peaks as the average fiber length decreases. Finally, the surface roughness map also indicates the relative position of the sample surface to the mathematical limit in the relationship between skewness and kurtosis according to the Pearson inequality - black dashed parabola in **Figure 13**, under which no mathematically sound Rsk, Rku pairs exist. In the case of the HEWL AF surfaces, the dialyzed HEWL AF sample resulted in the longest fibers and lowest kurtosis of all three samples (**Figure 13f**). Kurtosis approaching the Pearson inequality limit indicates a binomial distribution. In the context of surface height characteristics, this sample exhibits a height distribution dominated by two values. These surfaces are not naturally occurring and may have different wearing, wetting, or bonding properties. The results from this analysis indicate that fiber length control can be leveraged to tune surface roughness, specifically its kurtosis.

# 4. Conclusions

This work developed a Python-implemented analysis designed for characterizing periodic features of nanostructured materials. The analysis quantified the characteristic length of periodic features such as surface wrinkles on thin films and the average diameter of droplet populations entrained in emulsions. The open-source software developed leverages 2D-FFT image analysis for periodic feature detection and power spectral density analysis for characteristic signal identification. It was validated using synthetic images with known periodicity. The Python analysis was applied to biaxial and uniaxial nanowrinkled thin films of CNC-POEGMA and CNC-XG composites and CNC emulsion droplets. Additionally, this work used AFM image analysis to estimate surface roughness parameters of HEWL AF spin-coated surfaces.

The characteristic length results obtained provided wrinkle sizes with increased accuracy over manual measurements for all thin film surfaces considered in this work. Manual measurements of surface morphologies with more visually homogeneous features suffered from a relative error of 18.8% (biaxial) and 3.3% (uniaxial). Manual measurements of a visually heterogeneous biaxial morphology resulted in a considerably higher relative error of 51.2%. However, further analysis of this morphology's radially averaged power spectral density revealed a secondary frequency signal corresponding to the manually measured wrinkle size. This indicated a measurement bias in manually sampling a heterogeneous wrinkle morphology, which can be eliminated using the Python periodic feature analysis developed in this work. The physical "noise" in surface morphology was quantified using FWHM and PQF of the PSD's characteristic peak. The biaxial CNC-POEGMA sample was determined to have a higher FWHM (broader peak), translating to a more heterogeneous morphology compared to the uniaxial CNC-POEGMA sample. Furthermore, when comparing the two biaxially wrinkled surface samples with different compositions, the CNC-XG morphology had a smaller FWHM (narrower peak) corresponding to a relatively *less* heterogeneous morphology than the CNC-POEGMA surface.

This work also examined the effectiveness of the Python analysis on determining characteristic length for droplets entrained in a CNC emulsion. The analysis determined two droplet size populations with average diameters of 7.1 and 0.87 μm. The Python analysis was found to be particularly useful in the characterization of droplet sizes, as it was not possible to discern between population sizes through visual inspection in manual measurements. As a result, the average droplet diameter from manual measurements was 2.1 μm - a value between the two droplet population sizes above. This highlights the importance of a systematic, robust, and unbiased methodology, like the one presented here, to characterize a highly variable data set that may hide valuable information in its average.

The Python program presented in this work also implemented a surface roughness characterization. Definitions of surface roughness parameters were presented in terms of the third and fourth statistical moments of surface height profiles. AFM images of HEWL AFs were analyzed and compared to sample biological surfaces on a surface roughness map. It was found that the average fiber length of the AF suspension used had an inverse relationship with surface kurtosis (sharpness of surface peaks). In contrast, the skewness (bias) of the three surfaces tested was similar.

The scripts, data, and tools used for analysis in this work are available for open access, reference, and use in a GitHub repository[17]. This repository offers the opportunity for open-source use of the tools presented here and to develop new ones to strengthen this repository's applicability and extend its capabilities by adding new analyses.

# Author Contributions

# Acknowledgments

The authors acknowledge the support from the Digital Research Alliance of Canada, Calcul Quebec, and WestGrid through computational resource grants, expertise, and technical support. Cassandra Koitsopoulos and Lenka Vitkova are gratefully acknowledged for providing the CNC emulsion optical microscopy images and the lysozyme amyloid atomic force microscopy images, respectively.

# Funding

AG acknowledges financial support for the work presented here from the NSERC Canada Graduate Scholarship (CGS-D) and the McGill Graduate Mobility Travel Award. KD also acknowledges financial support from the Natural Sciences and Engineering Research Council of Canada (NSERC, Grant No. RGPIN-2023-03607).

# Declaration of Competing Interest

The authors declare that they have no known competing financial interests or personal relationships that could have appeared to influence the work reported in this paper.

# References

1. Cooley J, Tukey J. An algorithm for the machine calculation of complex Fourier series. Mathematics of Computation. 1965;19: 297–301.

2. De France K, Zeng Z, Wu T, Nyström G. Functional materials from nanocellulose: Utilizing structure-property relationships in bottom-up fabrication. Adv Mater. 2021;33: e2000657.

3. Verhulsel M, Vignes M, Descroix S, Malaquin L, Vignjevic DM, Viovy J-L. A review of microfabrication and hydrogel engineering for micro-organs on chips. Biomaterials. 2014;35: 1816–1832.

4. Tokarev I, Minko S. Stimuli-responsive hydrogel thin films. Soft Matter. 2009;5: 511–524.

5. White EM, Yatvin J, Grubbs JB III, Bilbrey JA, Locklin J. Advances in smart materials: Stimuli-responsive hydrogel thin films. J Polym Sci B Polym Phys. 2013;51: 1084–1099.

6. Yuk H, Zhang T, Parada GA, Liu X, Zhao X. Skin-inspired hydrogel-elastomer hybrids with robust interfaces and functional microstructures. Nat Commun. 2016;7: 12028.

7. Han L, Lu X, Wang M, Gan D, Deng W, Wang K, et al. A mussel-inspired conductive, self-adhesive, and self-healable tough hydrogel as cell stimulators and implantable bioelectronics. Small. 2017;13. doi:10.1002/smll.201601916

8. Stimpson TC, Wagner DL, Cranston ED, Moran-Mirabal J. Image analysis of structured surfaces for quantitative topographical characterization. ChemRxiv. 2020. doi:10.26434/chemrxiv.12736289.v1

9. Groenewold J. Wrinkling of plates coupled with soft elastic media. Physica A. 2001;298: 32–45.

10. De France KJ, Babi M, Vapaavuori J, Hoare T, Moran-Mirabal J, Cranston ED. 2.5D Hierarchical Structuring of Nanocomposite Hydrogel Films Containing Cellulose Nanocrystals. ACS Appl Mater Interfaces. 2019;11: 6325–6335.

11. Gill U, Sutherland T, Himbert S, Zhu Y, Rheinstädter MC, Cranston ED, et al. Beyond buckling: humidity-independent measurement of the mechanical properties of green nanobiocomposite films. Nanoscale. 2017;9: 7781–7790.

12. Wang Z, Servio P, Rey AD. Pattern formation, structure and functionalities of wrinkled liquid crystal surfaces: A soft matter biomimicry platform. Front Soft Matter. 2023;3: 1123324.

13. Busse A, Jelly TO. Effect of high skewness and kurtosis on turbulent channel flow over irregular rough walls. J Turbul. 2023;24: 57–81.

Pre-print
14. Usov I, Mezzenga R. FiberApp: An Open-Source Software for Tracking and Analyzing Polymers, Filaments, Biomacromolecules, and Fibrous Objects. Macromolecules. 2015;48: 1269–1280.

15. De France KJ, Xu F, Toufanian S, Chan KJW, Said S, Stimpson TC, et al. Multi-scale structuring of cell-instructive cellulose nanocrystal composite hydrogel sheets via sequential electrospinning and thermal wrinkling. Acta Biomater. 2021;128: 250–261.

16. Martínez Y, Heeb M, Kalač T, Gholam Z, Schwarze FWMR, Nyström G, et al. Biopolymer-based emulsions for the stabilization of Trichoderma atrobrunneum conidia for biological control. Appl Microbiol Biotechnol. 2023;107: 1465–1476.

17. Guerra A. morpholoPy: Python package for periodic features identification and characterization of nanostructured surfaces and emulsions. Github; Available: https://github.com/DReGuerra/morpholoPy

18. Photopea. [cited 18 Jan 2025]. Available: https://www.photopea.com/

19. Wang Z, Servio P, Rey AD. Cholesteric liquid crystal roughness models: from statistical characterization to inverse engineering. Soft Matter. 2025; Under Review.

20. Canny J. A computational approach to edge detection. IEEE Trans Pattern Anal Mach Intell. 1986;PAMI-8: 679–698.

21. Schindelin J, Arganda-Carreras I, Frise E, Kaynig V, Longair M, Pietzsch T, et al. Fiji: an open-source platform for biological-image analysis. Nat Methods. 2012;9: 676–682.

22. De France KJ, Kummer N, Ren Q, Campioni S, Nyström G. Assembly of Cellulose Nanocrystal–Lysozyme Composite Films with Varied Lysozyme Morphology. Biomacromolecules. 2020;21: 5139–5147.

23. Kummer N, Wu T, De France KJ, Zuber F, Ren Q, Fischer P, et al. Self-assembly pathways and antimicrobial properties of lysozyme in different aggregation states. Biomacromolecules. 2021;22: 4327–4336.

24. Wainwright DK, Lauder GV, Weaver JC. Imaging biological surface topography in situ and in vivo. Methods in Ecology and Evolution. 2017;8: 1626–1638.

# Supplemental Information

The Python code used for synthetic and experimental image analyses presented in this work can be found in a GitHub repository[17]. This repository will be continuously developed to improve the available scripts and add new nanostructured surface image analysis tools.

# Additional Test Cases - Synthetic Patterns

The following three test cases followed the same two-dimensional FFT analysis described in this work's Methods and Validation section. This supplemental section briefly highlights the differences between the test cases below and the one described in the main text. These test cases present images with increasingly more complex patterns to progressively introduce more randomness to the image.

## Chevron

The chevron pattern was generated from the image of the vertical lines described in the main text. This test case is in the GitHub repository referenced above in the `test_cases/synthetic_images/chevron/` directory. The image was generated by periodically shifting the pixels left and right to create 45-degree angles in the black and white stripes.

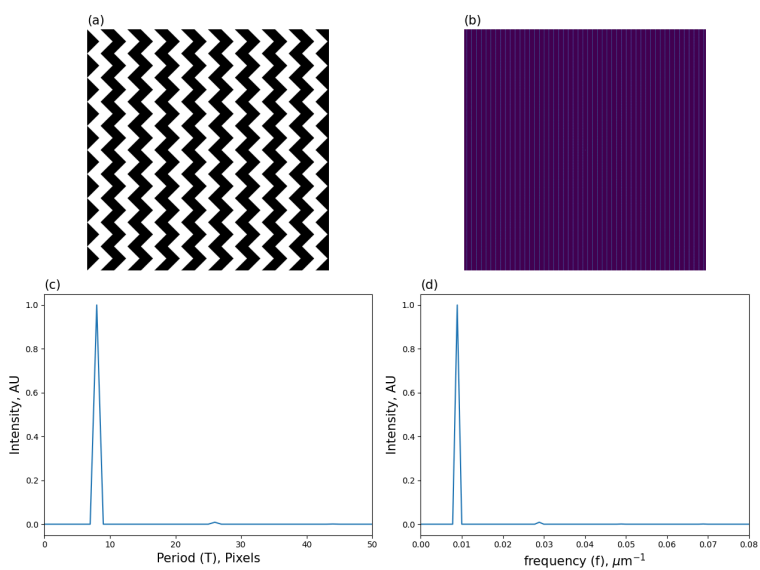

**Figure S1: Chevron pattern.** (a) The 900x900 pixel 8-bit image of a black and white chevron pattern, (b) the natural log of 1 plus magnitude of the center-shifted 2D-FFT of the image (i.e., 2D-PSD) of the image), and (c) the radially averaged 2D-PSD of the image in the pixel frequency domain, and (d) the radially averaged 2D-PSD of the image in the spatial frequency domain.

# Additional Test Cases - Experimental Systems

## CNC Emulsion Droplets

This section contains two more images captured from the CNC emulsion system analyzed using the same procedure described in the main text. **Figures S2** and **S3** reveal characteristic lengths (droplet diameters) that agree with each other, and the first sample image is presented in **Figure 11**. This was found for both the largest and the second-largest populations of droplets.

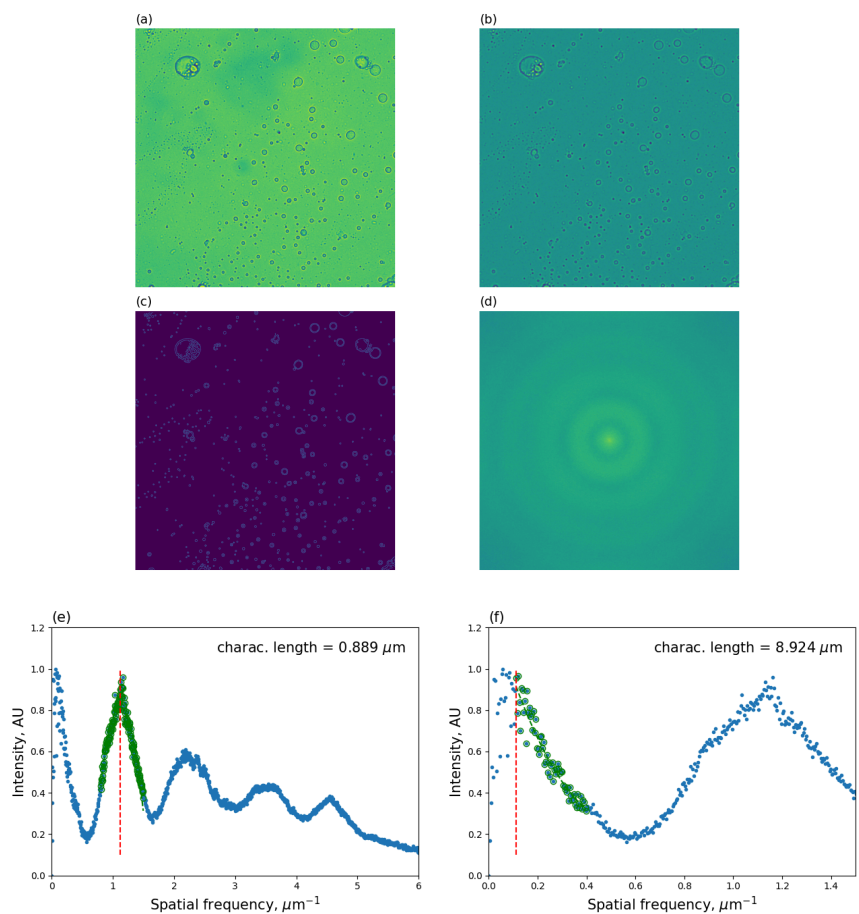

**Figure S2: CNC emulsion with various-sized droplets entrained.** (a) The original image of the CNC emulsion with droplets, (b) low band-pass ($\sigma_{low} = 3.0$) and high-band pass ($\sigma_{high} = 12$) filtered image by a difference of Gaussian, (c) edge detected image using the Canny algorithm ($\sigma_{Canny} = 1.4$), (d) 2D-PSD of the edge-detected image, (e and f) the radially averaged power spectral density with quadratic regression applied to the two most prominent peaks in the scale of interest to identify characteristic length of periodic features of the surface (droplet diameters) identified by a dashed red vertical line.

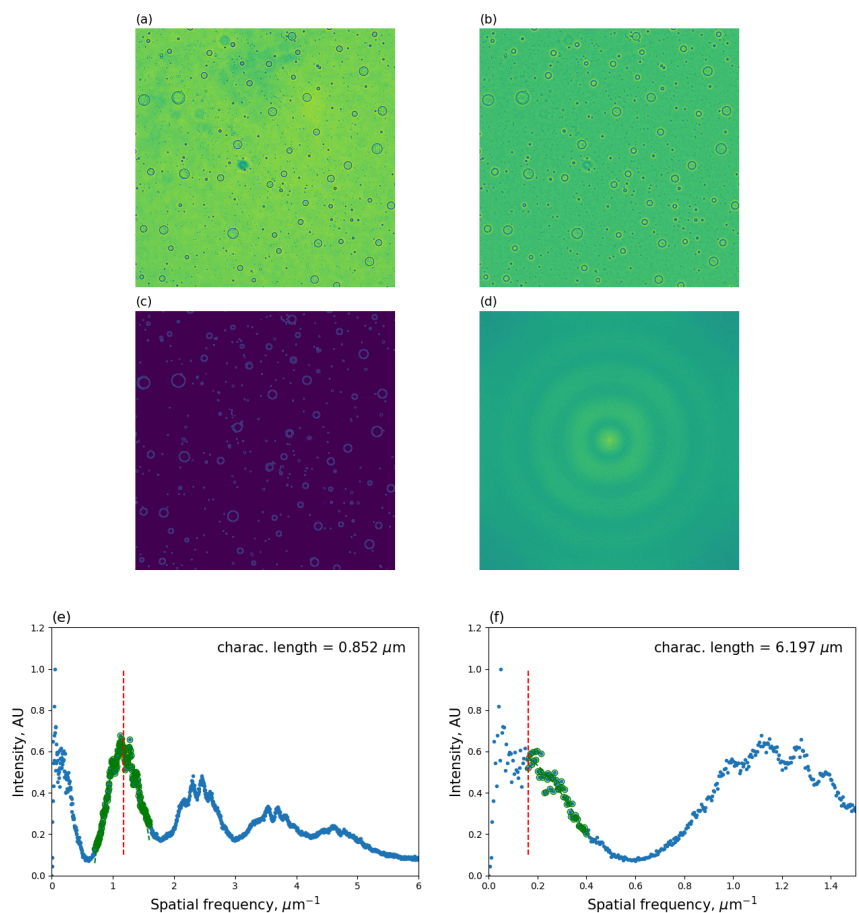

**Figure S3: CNC emulsion with various-sized droplets entrained.** (a) The original image of the CNC emulsion with droplets, (b) low band-pass ($\sigma_{low} = 3.0$) and high-band pass ($\sigma_{high} = 12$) filtered image by a difference of Gaussian, (c) edge detected image using the Canny algorithm ($\sigma_{Canny} = 1.4$), (d) 2D-PSD of the edge-detected image, and (e and f) the radially averaged power spectral density with quadratic regression applied to the two most prominent peaks in the scale of interest to identify characteristic length of periodic features of the surface (droplet diameters) identified by a dashed red vertical line.